\documentclass[USenglish,twocolumn]{article}

\usepackage[utf8]{inputenc}
\usepackage[big]{dgruyter}
\usepackage{microtype}
\usepackage{color,soul}
\usepackage{csquotes}
\usepackage{braket}

\usepackage[doi=false,url=false, isbn=false,backend=bibtex,style=phys,chaptertitle=true,biblabel=brackets,articletitle=true,maxbibnames=5,minbibnames=3,pageranges=false,defernumbers=true]{biblatex}
\usepackage{jabbrv}
\DeclareFieldInputHandler{journaltitle}{%
  \def\NewValue{\JournalTitle{#1}}}
\DeclareBibliographyCategory{cited}
\AtEveryCitekey{\addtocategory{cited}{\thefield{entrykey}}}

\usepackage{xcolor}
{\catcode `\@=11 \global\let\AddToReset=\@addtoreset}
\AddToReset{equation}{section}
\renewcommand{\theequation}{\thesection.\arabic{equation}}
\newcounter{mnotecount}[section]
\renewcommand{\themnotecount}{\thesection.\arabic{mnotecount}}
\newcommand{\mnotex}[1]
{\protect{\stepcounter{mnotecount}}$^{\mbox{\footnotesize
$
\bullet$\themnotecount}}$ \marginpar{
\raggedright\tiny\em
$\!\!\!\!\!\!\,\bullet$\themnotecount: #1} }

\renewcommand{\hl}[1]{{\color{black} #1}}

\begin{document}

  \articletype{Views}

  \author[1]{Markus Rademacher}
  \author[2]{James Millen}
  \author*[3]{Ying Lia Li} 
  \runningauthor{...}
  \affil[3]{Department of Physics \& Astronomy, University College London, London WC1E 6BT, UK, e-mail: ying.li.11@ucl.ac.uk}
  \affil[1]{Department of Physics \& Astronomy, University College London, London WC1E 6BT, UK}
  \affil[2]{Department of Physics, King’s College London, Strand, London, WC2R 2LS, UK}
  \title{Quantum sensing with nanoparticles for gravimetry: when bigger is better}
  \runningtitle{...}
  \subtitle{...}
  \abstract{Following the first demonstration of a levitated nanosphere cooled to the quantum ground state in 2020~\cite{delic_cooling_2020}, macroscopic quantum sensors are seemingly on the horizon. The nanosphere's large mass as compared to other quantum systems enhances the susceptibility of the nanoparticle to gravitational and inertial forces. In this viewpoint we describe the features of experiments with optically levitated nanoparticles \cite{millen_optomechanics_2020} and their proposed utility for acceleration sensing. Unique to the levitated nanoparticle platform is the ability to implement not only quantum noise limited transduction, predicted by quantum metrology to reach sensitivities on the order of $10^{-15}$\,ms$^{-2}$\hl{{~\cite{Qvarfort2018}},} but also long-lived quantum spatial superpositions for enhanced gravimetry. This follows a global trend in developing sensors, such as cold atom interferometers, that exploit superposition or entanglement. Thanks to significant commercial development of these existing quantum technologies, we discuss the feasibility of translating levitated nanoparticle research into applications.  
}
  \communicated{...}
  \dedication{...}
  \received{...}
  \accepted{...}
  \journalname{...}
  \journalyear{...}
  \journalvolume{..}
  \journalissue{..}
  \startpage{1}
  \aop
  \DOI{...}

\maketitle

\section{Introduction}
Quantum mechanics is a cornerstone of modern physics, and quantum behaviour, such as superposition and entanglement, have been extensively observed using subatomic particles, photons, and atoms since the early 1900's~\cite{davisson_reflection_1928}. A global effort is now underway to take these existing quantum experiments and devices out of the laboratory and into industry, in a process dubbed the `second quantum revolution'.

\begin{figure*}[!ht]
    \centering
    \includegraphics[width=1.5\columnwidth]{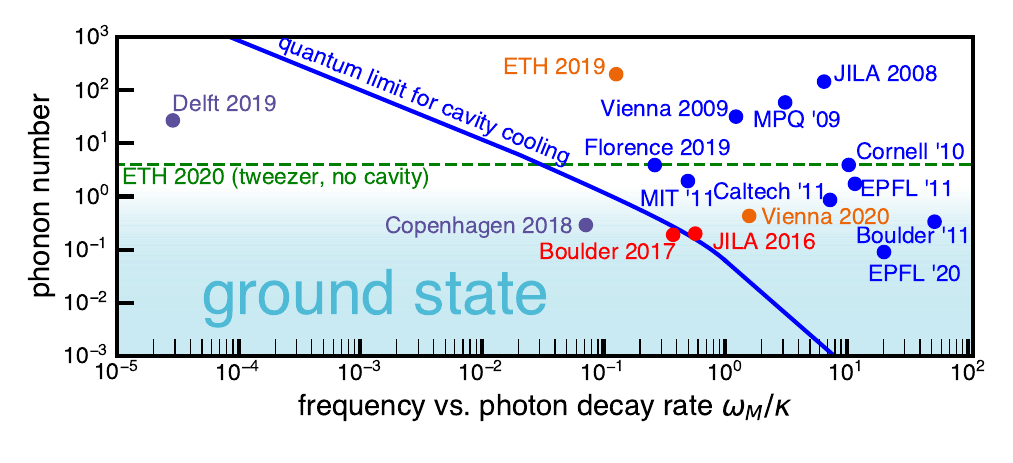}
    \caption{Experimental results for cooling of macroscopic systems. Minimum phonon occupation number is plotted against the sideband-resolvability parameter $\omega_m/\kappa$. Blue solid line displays the minimum achievable phonon number using quantum limited passive cavity cooling. Blue data points represent experiments relying only on passive cavity cooling. Red data points are results using squeezed light to surpass the standard quantum limit imposed on cavity cooling. Purple data points present results using a feedback cooling scheme. Orange data points show recent results of cooling levitated nanoparticles using coherent scattering in a cavity. Green dashed line shows recent data of a nanoparticle feedback cooled in an optical tweezer using no cavity for cooling or read out purposes. EPFL '20:~\cite{qiu_laser_2020}; Vienna 2020:~\cite{delic_cooling_2020}; ETH 2020:~\cite{tebbenjohanns_motional_2020}; ETH 2019:~\cite{windey_cavity-based_2019}; Delft 2019:~\cite{guo_feedback_2019}; Florence 2019:~\cite{chowdhury_calibrated_2019}; Copenhagen 2018:~\cite{rossi_measurement-based_2018}; Boulder 2017:~\cite{clark_sideband_2017}; JILA 2016:~\cite{peterson_laser_2016}; Boulder '11:~\cite{teufel_sideband_2011}; Caltech '11:~\cite{chan_laser_2011}; EPFL '11:~\cite{riviere_optomechanical_2011}; MIT '11:~\cite{schleier-smith_optomechanical_2011}; Cornell '10:~\cite{oconnell_quantum_2010}; MPQ '09:~\cite{schliesser_resolved-sideband_2009}; Vienna 2009:~\cite{groblacher_demonstration_2009}; JILA 2008:~\cite{teufel_dynamical_2008};}
    \label{fig:quantumoptomechanics}
\end{figure*}

Technological advances are now enabling larger objects to enter the quantum regime, with 2010 heralding the first ground state cooling of the motion of a human-made object, \hl{specifically a micron-scale `quantum drum'} \cite{oconnell_quantum_2010}.  Operating in the quantum regime with \emph{free} or \emph{levitated} particles would allow the generation of macroscopic quantum states, and enable greatly enhanced sensitivity to external forces. The state-of-the-art demonstration of a macroscopic superposition is currently provided by matter-wave interferometry with an engineered molecule of mass beyond 25,000\,Da \cite{fein_quantum_2019}. This year, the centre-of-mass (c.o.m.) motion of a 143\,nm diameter silica nanosphere, levitated within an optical cavity, was cooled to its zero point energy (average phonon occupancy $<1$) using the cavity optomechanical interaction \cite{delic_cooling_2020}. Significant developments in trapping, stabilisation and cooling techniques (see Section~\ref{implementation}) have enabled levitated systems to reach the quantum regime (as shown in Figure~\ref{fig:quantumoptomechanics}), bringing researchers closer towards generating macroscopic quantum states with solid nanoscale objects.

A driving force for creating quantum states of more massive objects, beyond proving their feasibility, is to test theories of gravity. The gravitational interaction has so far presented itself as classical \cite{carney_tabletop_2019}. It is unknown whether gravity acts as a quantum interaction, \hl{for example, via virtual graviton exchange~{\cite{marshman2020}},} or if in fact gravity is responsible for wave function collapse. In the latter case, one extends the Schr\"odinger equation nonlinearly to account for gravitational self-interaction, formalised in the Schr\"odinger-Newton equation\hl{. Models of gravitationally induced wave function collapse} aim to define the timescale of collapse due to a superposition of two different space-time curvatures, whilst avoiding superluminal signalling \cite{penrose_gravitys_1996,penrose_gravitization_2014}. Nonlinear modifications to the Schr\"odinger equation are also studied in the continuous spontaneous localisation (CSL) model~\cite{ghirardi_markov_1990}\hl{. This} aims to justify quantum wave-function collapse by introducing a stochastic diffusion process driven by an unknown noise field that continuously counteracts the spread of the quantum wave function. \hl{Through experiment, these collapse models can be falsified to rule out a mass-limit on quantum superpositions due to gravitational or noise induced localization} ~\cite{bassi_models_2013,martin_cosmic_2020, nimmrichter_optomechanical_2014}. The CSL effect would be practically unobservable on the atomic level but strongly amplified for high-mass systems. Many collapse models can be discounted by the sheer act of observing matter-wave interference with increasingly massive objects, such as levitated nanoparticles~\cite{bateman_near-field_2014,arndt_testing_2014}.

For sensing applications, dense macroscopic systems offer an enhanced sensitivity to acceleration, such that a single quantum nanosphere is predicted to reach acceleration sensitivities $10^{5}$ times more sensitive than a cloud of cold atoms \hl{{\cite{Qvarfort2018}}}. If used for navigation applications, this improvement in sensitivity reduces the  accumulated error in position caused by double integrating a less noisy acceleration signal. Similar to cold atoms, levitated nanospheres are \hl{well} isolated from environmental decoherence, resulting in long coherence times for matter-wave interferometry and the ability to perform free-fall experiments. Free-fall accelerometers are particularly suited for gravimetry applications aimed at resolving the temporal and spatial fluctuations of gravitational acceleration at the Earth's surface, which can vary roughly between 9.78\,ms$^{-2}$ and 9.83\,ms$^{-2}$~\cite{menoret_gravity_2018}. Through gravimetry, one can directly infer information about sub-surface mass distributions, including volcanic activity monitoring~\cite{carbone_added_2017}, ice mass changes~\cite{makinen_absolute_2007}, subsidence monitoring~\cite{van_camp_repeated_2011}, and the detection of underground cavities~\cite{romaides_comparison_2001}. The latter is of interest to the oil and gas industry, as well as the construction industry. Section~\ref{sec:sense} lists a range of quantum sensing proposals involving levitated nanospheres suitable for these types of applications.

\section{Implementation} 
\label{implementation}
Macroscopic mechanical oscillators which are controlled using light belong to a field of study called optomechanics. The use of a resonant optical cavity can significantly enhance the light-matter interaction. At the heart of all cavity optomechanical systems is a dispersive interaction, where an optical resonance frequency is shifted due to mechanical motion. This governs the read out of zero-point fluctuations and any motion caused by forces acting on the system. Before explaining the benefits of such a sensing scheme, we first describe the main components of an optomechanical system, and the variety of mechanical modes and optical resonances employed by researchers. 

\subsection{Typical features of cavity optomechanics}
\begin{figure}[!ht]
    \centering
    \includegraphics[width=\columnwidth]{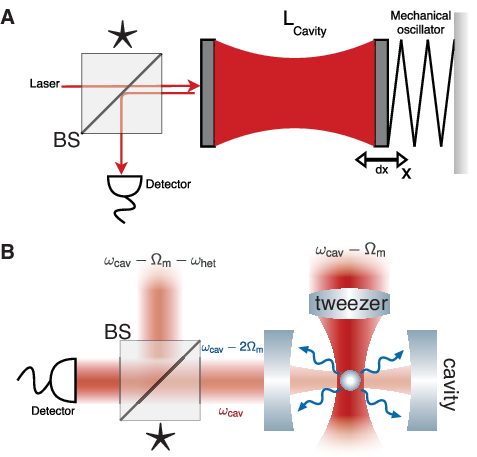}
    \caption{(A) Basic set-up of an optomechanical interferometer formed by coupling an optical mode to the mechanical motion of the end-mirror of a Fabry-Perot cavity. (B) Schematic set-up for cavity cooling of a levitated nanoparticle using the coherent scattering technique, as demonstrated in~\cite{delic_cooling_2020}, where the cavity is not directly pumped as it is in (A). \hl{The star in (A) and (B) indicates the position of an additional Fabry-Perot cavity required to create superposition states, as explained in section~\ref{sec:optical-superposition}.}}
    \label{fig:schematic}
\end{figure}

In general, a cavity optomechanical system consists of three main ingredients. Firstly, a  mechanical mode, such as the centre-of-mass (c.o.m.) motion of the end-mirror of a Fabry-Perot cavity, as shown in Figure~\ref{fig:schematic}(A). It can also be the c.o.m. motion of a levitated nanoparticle \cite{chang_cavity_2010,kiesel_cavity_2013}, a membrane \cite{chan_laser_2011} or a cantilever structure \cite{metzger_cavity_2004}. Levitated (nanoparticle) optomechanics benefits from \hl{excellent} environmental isolation \cite{millen_optomechanics_2020} whereas some clamped systems possess exceptionally high frequency mechanical modes in the microwave (MW) frequency range~\cite{kippenberg_cavity_2008}. Typical c.o.m. oscillation frequencies of levitated nanoparticles range from a few \,kHz to several hundred \,kHz~\cite{kiesel_cavity_2013,delic_cooling_2020,millen_optomechanics_2020}. The mass of nanoparticles typically employed in levitated optomechanics varies from $10^{-19}$\,kg to $10^{-16}$\,kg~\cite{delic_cooling_2020,millen_cavity_2015,millen_optomechanics_2020}.~\footnote{For levitated systems, rotational and librational motion at higher frequencies is also studied \cite{kuhn_optically_2017,ahn_optically_2018,reimann_ghz_2018,rashid_precession_2018,rahman_laser_2017}. For clamped systems, internal mechanical modes such as radial breathing modes, which can possess GHz frequencies, are routinely used to demonstrate near-quantum ground state preparation \cite{schliesser_resolved-sideband_2009}. Higher order mechanical modes are also studied \cite{jiao_nonlinear_2016}.}

The second requirement is a cavity-confined optical mode which is coupled to the mechanical oscillator. Figure~\ref{fig:schematic}(A) illustrates a resonant standing wave within a Fabry-Perot cavity, which can be coupled to the motion of the cavity end-mirror via radiation pressure. Alternative optical modes include the evanescent fields of whispering gallery mode \cite{schliesser_cavity_2014} and photonic crystal \cite{eichenfield_optomechanical_2009,magrini_near-field_2018} resonances, which can be coupled to their own internal mechanical modes or to external mechanical oscillators. A narrow cavity resonance linewidth $\kappa$ enables one to reach the `resolved-sideband regime', where the mechanical oscillation frequency $\Omega_{m}$ is larger than $\kappa$. This allows energy transfer between the optical and mechanical modes in an anti-Stokes/Stokes process, enabling cooling of the mechanical oscillator. A cavity is not necessarily required to reach the ground state \cite{tebbenjohanns_motional_2020}, but a cavity provides resonant enhancement in read out and interaction strength\hl{. This reduces} the number of photons needed to interact with the mechanical oscillator, improving the signal to noise. 

Thirdly, a sufficiently high optomechanical coupling rate $G$ (in units of Hz/m) is required. This encodes the shift in the optical cavity resonance frequency caused by the motion of the mechanical oscillator. Large $G$ is required to optically transduce, control or cool the oscillator, with the highest $G$ obtained when the overlap between the optical field and the displacement field is maximised \cite{camacho_characterization_2009}, i.e. by using a spatially confined optical mode.

\subsection{Ground state cooling}
In this viewpoint, we focus on quantum enhanced sensing using levitated systems. Below, we explain the coherent scattering technique, and how it enables cooling of the c.o.m. mode of a macroscopic object to the ground state of an optical potential \cite{delic_cooling_2020}.

 A macroscopic quantum state can be created by cooling the c.o.m. motion of a nanoparticle levitated within a harmonic potential, with a mechanical frequency $\Omega_{m}$. The position uncertainty of the particle is $\sigma_x = \sqrt{\hbar(1+2n)/2m\Omega_m}$, where the phonon occupancy $n$ is related to the c.o.m. temperature $T_{\rm{CM}}$ through $n = \sqrt{k_BT_{\rm{CM}}/\hbar\Omega_m}$. When cooled to the ground state, the particle has a position uncertainty, or zero-point fluctuation, of $\sigma_{\rm{zpf}} = \sqrt{\hbar/2m\Omega_m}$. If the particle is released from the levitating potential, this position spread \hl{grows approximately linearly in time{~\cite{romero-isart_coherent_2017}}}. Considering typical parameters for a levitated nanoparticle of $m = 10^{-18}\,$kg and $\Omega_m = 2\pi\times 10^5\,$rad/s, this yields $\sigma_{\rm{zpf}} \approx 10^{-11}\,$m, requiring hundreds of seconds of expansion until the quantum position spread is as large as the particle, a reasonable definition of a macroscopic quantum state. However, subsequent matter-wave interferometry can be used to boost the size of the quantum state~\cite{millen_optomechanics_2020}, discussed further in Section~\ref{sec:sense}.

Following the above discussion, the particle must be initially cooled near to the ground state of the levitating potential. A range of passive and active cooling methods to achieve this are described in multiple review papers \cite{aspelmeyer_cavity_2014}, with many techniques such as sideband resolved cooling derived from the cold atom community \cite{ritsch_cold_2013}. Here we focus on the `coherent scattering' protocol \cite{vuletic_laser_2000,delic_cavity_2019,windey_cavity-based_2019}, as illustrated in Figure~\ref{fig:schematic}(B), which is the most successful cooling technique for levitated nanoparticles.

The mechanical oscillator is the c.o.m. motion of a levitated nanoparticle, and it is held within an optical cavity using an optical tweezer, as shown in Figure~\ref{fig:schematic}(B). The frequency of the oscillator is set by the optical potential provided by this single-beam gradient force trap. The tweezer allows the optimal placement of the particle within the cavity mode;  
the coupling strength is at its highest when the particle is held at the cavity node. 

The optical cavity is not pumped externally. The trapping optical tweezer frequency is stabilized relative to the cavity resonance, and light scattered out of the tweezer field by the nanosphere then populates the cavity mode, interacting coherently with the oscillator again. It is a key feature of the coherent scattering technique that the optical cavity is only pumped by the light scattered by the nanoparticle. 

Consequently, each photon populating the cavity mode interacts with the particle, increasing the optomechanical coupling rate. As a result, the quantum cooperativity \footnote{The cooperativity is defined as a ratio of the optomechanical coupling strength and the product of the optical and mechanical decay rates.} of the experiment is well above 1000. To put this into perspective, a quantum cooperativity >1 is the benchmark for entering the quantum back-action regime~\cite{millen_optomechanics_2020}; a long sought-after goal in levitated optomechanics. A high cooperativity is also known in cold atom physics to produce a constant cooling rate for cavity assisted molecule cooling in dynamical potentials~\cite{ritsch_cold_2013,chang_quantum_2014}. Compared to externally pumped cavity cooling schemes~\cite{delic_levitated_2020}, the estimated improvement in cooperativity is $10^5$-fold~\cite{delic_thesis_2019} due to coherent scattering.

One of the biggest challenges that previously prevented ground state cooling is a method to circumvent heating due to scattering and phase noise in the optical cavity. Phase noise in the cavity field can be reduced by increasing optical power, with the trade-off that scattering noise increases. In the coherent scattering scheme, laser phase noise is almost completely evaded since optimal cooling of the nanoparticle occurs at the cavity node, where the intensity minimum of the cavity standing wave is located. Further noise reduction involves a balance between increasing the optomechanical coupling strength and decreasing the scattering noise, whilst still ensuring the particle is stably levitated.

Finally, the coherent scattering cooling scheme is inherently multidimensional. Although to date only one axis has been optimized for ground state cooling, when rotating the trap accordingly full 3D cooling is possible~\cite{windey_cavity-based_2019}. Varying the coupling of each c.o.m. degree of freedom to the cavity by moving and tuning the scattering plane of the optical tweezer makes the coherent scattering implementation flexible. In contrast, only strong one dimensional cooling can be achieved in cavity systems such as Figure~\ref{fig:schematic}(A), where the static intracavity trapping potential limits cooling to be along the cavity axis. 

Recent theoretical work on how to treat the multidimensional cooling dynamics illustrate apparent 3D hybridisation effects in coherent scattering. These hybridising pathways act as a road map to engineer displacement sensing possibly surpassing the standard quantum limit (SQL)~\cite{toros_quantum_2020}.

\section{Inertial Sensing \& Gravimetry}
\label{sec:sense}
Macroscopic quantum objects offer significant sensing advantages over their lower-mass cold atom counterparts through an enhanced coupling to inertial and gravitational forces\hl{\footnote{\hl{This is generally true for inertial and relative gravity sensors based on mechanical oscillators. For free-fall absolute measurements of gravity, the mass component often cancels out due to the equivalence principle.}}}. Generally, the competitive edge of optomechanical sensors can be summarised through two sensing strategies; quantum limited transduction, and sensing which exploits either superposition or entanglement, the features of the so-called `second quantum revolution'. Levitation \hl{provides excellent} environmental isolation, ensuring long coherence times in which to perform sensitive measurements. For example, the coherent scattering experiment mentioned above achieves a coherence time of 7.6\,$\mu s$, corresponding to 15 coherent oscillations before the ground state is populated by even a single phonon~\cite{delic_cooling_2020}. 

We will first discuss the limitations of continuous optical transduction of the oscillator position, before exploring the use of spatial superpositions to extract a gravitationally induced phase shift via matter-wave interferometry. We will describe proposals that increase this phase shift through coupling to spin, creating a spin-oscillator superposition where gravity or acceleration is read out using the spin state.

\subsection{Continuous optical sensing}

The dispersive interaction between a mechanical oscillator and an optical resonance allows for continuous read out of the oscillator motion through probing the optical field quadrature. When thermal motion is present, the acceleration sensitivity for frequencies below mechanical resonance remains bound by $a_{\rm{th}}=\sqrt{\frac{4k_{B}T_{\rm{CM}}\Omega_{\rm{m}}}{m Q_{\rm{m}}}}$, where $T_{\rm{CM}}$ is the cooled mode temperature, $\Omega_{\rm{m}}$ the oscillator frequency and $Q_{\rm{m}}$ the mechanical quality factor. During cooling, $T_{\rm{CM}}$ decreases proportionally with $Q_{\rm{m}}$, resulting in no net change to the noise floor $a_{\rm{th}}$ \cite{krause_high-resolution_2012}. However, cooling does reduce the classical thermomechanical noise at the mechanical frequency. With the mechanical oscillator prepared in the ground state, an optomechanical measurement of the position is no longer limited by thermal motion, but by the zero point fluctuations and added noise from the transduction (measurement). The sensing sensitivity is now set by the SQL where back-action noise caused by photon momentum kicks, and imprecision noise caused by phase fluctuations contribute equally, yielding a displacement read out sensitivity of $2\times \sigma_{\rm{zpf}}$. 

However, the SQL does not correspond to a fundamental quantum limit \cite{mason_continuous_2019}. Methods to read out the mechanical motion whilst minimizing or evading the effects of back-action are known as quantum nondemolition measurements\hl{.  In these cases, the measured observable commutes with the system Hamiltonian, as provided by} coupling to the velocity of a free mass \cite{purdue_practical_2002} or modification of the mechanical susceptibility through engineering an optical spring to establish a new SQL \cite{arcizet_beating_2006}. The SQL can also be surpassed by monitoring only one of the two non-commuting quadratures of the motion, known as a back-action-evading measurement\hl{. This} can squeeze either the optical or mechanical quadrature, and is achieved using pulsed optomechanics \cite{vanner_pulsed_2011}.

\subsection{Spatial superpositions for sensing}
Superposition and entanglement are quantum effects, which have no classical analogue. Entanglement enables \hl{the distribution of quantum states between oscillators separated by a distance}, whilst superposition enables the oscillator to be in a linear sum of several motional states. When exploiting these effects for inertial sensing and gravimetry, additional sensitivity or resolution can be achieved. For example, multiple entangled quantum nanospheres could be used for distributed sensing. Such sensing schemes require additional experimental steps beyond cooling to prepare the quantum state.  In this section we describe a select number of quantum sensing proposals involving levitated nanoparticles, which are now within grasp. 

An alternative to performing continuous transduction of the nanoparticle's motion would be to prepare the particle in a spatial superposition and then perform matter-wave interferometry. To sense gravity would require the creation of a coherent superposition, such that the superposition is localized at two heights or regions of a varying local gravitational field. The two amplitudes of the wave-function evolve under the Newtonian gravitational potential, resulting in a relative phase difference which is then measured interferometrically. The phase difference is defined:
\begin{equation}
    \Delta\phi=\frac{1}{\hbar}\int_{0}^{t}\Delta U dt=\int_{0}^{t}\frac{mg\Delta z(t)\cos(\theta)}{\hbar}dt,
\end{equation}
where $\Delta U$ is the gravitational potential energy difference across a vertical spatial separation $\Delta z(t)$,  $m$ is the mass of the oscillator, and $\theta$ the angle between the interferometer and the direction of acceleration $g$ (here, defined by local gravity). The integral $\int^{t}_{0}\Delta z(t)dt$ is the path difference between the trajectories of the two amplitudes of the wave-function. For a trapped system, discussed in Section~\ref{sec:levsense}, $\Delta z(t)$ is the maximum spatial superposition separation, limited by the period of the trap oscillation $2\pi/\Omega_{\rm{m}}$. For an oscillator under free-evolution, discussed in Section~\ref{sec:freefall}, the particle wave-function freely evolves for as long as it remains coherent, enabling the creation of larger spatial superpositions~\cite{brawley_nonlinear_2016}. 

\subsubsection{Optically preparing spatial superpositions}
\label{sec:optical-superposition}
Generation of a quantum superposition requires a nonlinear interaction. To prepare a spatial superposition in an optical cavity a strong quadratic coupling to motion governed by $g_{2}~\hat{x}^{2}$, or a strong single-photon coupling utilising the nonlinearity of the radiation-pressure interaction, is required~\cite{brawley_nonlinear_2016}. For example, by using a laser pulse interaction that measures $\hat{x}^{2}$ via a homodyne measurement. This provides information of the nanosphere position relative to the cavity centre, but not the offset direction (left or right), creating a spatial superposition similar to a double slit. Technically, it is challenging to create either a strong linear or a strong quadratic optomechanical coupling.

Alternatively, superposition can be generated by entangling a photon with a modified optomechanical Michelson interferometer formed by adding an additional Fabry-Perot cavity at the unused port of the beamsplitter \hl{annotated with a star} in Figures~\ref{fig:schematic}(A,B). A superposition of the two optical cavity modes is formed by the beamsplitter interaction such that, without measurement, the photon enters both cavities at the same time. The radiation pressure of this photon causes a deflection of the mechanical oscillator of approximately the zero-point motion, thus creating a mechanical superposition where the mechanical oscillator is unperturbed and perturbed \cite{bose_scheme_1999}. 

\subsubsection{Spatial superpositions through coupling to spin}
\label{sec:levsense}
A nonlinear interaction can also be mediated by a two-level-system. In contrast to measuring the nanoparticle position through the phase modulation of the optical cavity field, or through generation of matter-wave interferometry, coupling the motion to a two-level-system enables read out through the two level states. The advantage of this method, which relies on discrete variables such as spin, is an immunity to motional noise which relaxes the need for ground state preparation. For example, many levitated nanosphere proposals that utilise spin only require a thermal state with moderate cooling. In general, because discrete variables allow the use of heralded probabilistic protocols, they also benefit from high fidelity and resilience to background noise or detection \hl{losses}.  

\begin{figure}[!ht]
    \centering
    \includegraphics[width=\columnwidth]{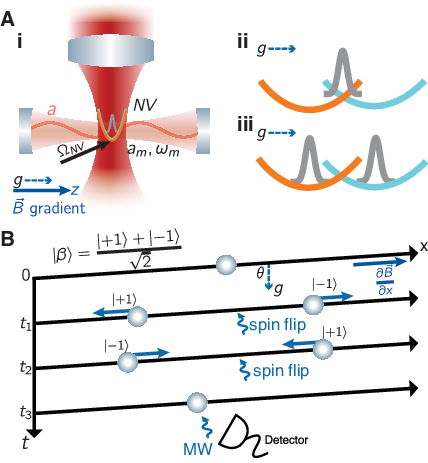}
    \caption{Spin-oscillator coupling has been proposed as a gravimetry technique, whereby a spatial superposition is created through the interaction of an embedded two-level-system such as a nitrogen-vacancy (NV) centre with an external magnetic field gradient. Variations of (A) has been proposed in~\cite{yin_large_2013,scala_matter-wave_2013,chen_high-precision_2018} A microwave pulse can be applied to split the spin states of the internal NV center in a trapped quantum nanosphere (i), which in turn creates a spatial superposition that can be viewed as the splitting of the optical trap (ii) into a superposition (iii). Note that the cavity is used to prepare the nanosphere in the ground state and can be switched off during (ii and iii). (B) Allowing for free evolution increases the spatial size of the superposition, created and probed in a Ramsey-type interferometer, as proposed by~\cite{wan_free_2016}. Here, the coupled NV-nanosphere superposition is prepared with a microwave (MW) pulse at time $t_{1}$, undergoing free-fall until the spins are flipped at $t_{2}$ to enable matter-wave interferometry at $t_{3}$.}
    \label{fig:gravimetry}
\end{figure}

Proposals combining levitated particles with two level systems include levitated nanodiamonds with an embedded nitrogen-vacancy (NV) centre with an electron spin \hl{{~\cite{yin_large_2013,scala_matter-wave_2013,chen_high-precision_2018,bose_spin_2017,marletto_gravitationally_2017}},} and a superconducting ring resonator coupled to a qubit \cite{johnsson_macroscopic_2016}. Here, we consider stationary spatial superpositions of a levitated nanoparticle oscillator with embedded spin, which remains trapped by an optical tweezer throughout the sensing protocol, as illustrated in Figure~\ref{fig:gravimetry}A. The spin is manipulated by microwave pulses. The first pulse introduces Rabi oscillations between the spin eigenvalue states $S_{z}=+1$ and $S_{z}=-1$, such that when a magnetic field gradient is applied the oscillator wavepacket is delocalized. This spin-dependent spatial shift is given by $\pm\Delta z=\frac{g_{\rm{nv}}\mu_{\rm{B}} B_{\rm{z}}}{2m\Omega^{2}}$, where $B_{\rm{z}}$ is the magnetic field gradient along the z-direction, which is the same direction that gravity acts in~\cite{chen_high-precision_2018,johnsson_macroscopic_2016}\footnote{A spatial superposition can be prepared at an angle $\theta$ to the acceleration force by tilting the applied magnetic field direction~\cite{scala_matter-wave_2013}}, $g_{\rm{nv}}\approx 2$ is the Land{\'e} $g$ factor and $\mu_{B}$ is the Bohr magneton. This effectively splits the harmonic trapping potential, creating a spatial superposition with equilibrium positions governed by a spin-dependent acceleration. The spin-oscillator system now has states $\ket{+1}$ and $\ket{-1}$ in different gravitational potentials, accumulating a relative gravitational phase difference. A measurement at $t_{0}=\frac{2\pi}{\Omega_m}$ yields a phase shift \cite{scala_matter-wave_2013}:
\begin{equation}
    \Delta\phi=\frac{16 \Lambda\Delta\lambda t_{0} }{\hbar^{2}\Omega_{m}}
\end{equation}
where $\Delta\lambda=\frac{1}{2}mg\cos(\theta)\sigma_{\rm{zpf}}$ is the gravity induced displacement, $\theta$ the angle between the applied magnetic field gradient $B_{\rm{z}}$ and the direction the nanosphere is accelerated (defined here as the direction of local gravity $g$), and $\sigma_{\rm{zpf}}$ the zero-point fluctuations. The spin-oscillator coupling is given by $\Lambda=g_{\rm{nv}}\mu_{\rm{B}}B_{\rm{z}}\sigma_{\rm{zpf}}$, with $g_{\rm{nv}}$ and $\mu_{\rm{B}}$ defined previously. This phase difference can be measured by applying a microwave pulse that probes the population of the spin state $S_{z}=0$ via $P_{0}=\cos^{2}(\frac{\Delta\phi}{2})$.

The maximum spatial superposition~\footnote{The path trajectory of each wavepacket is given by $z_{\pm}(t)=\pm\Delta z(1-\cos(\Omega_{\rm{m}}t))+\frac{g}{\Omega_{\rm{m}}^{2}}$, where $\frac{g}{\Omega_{\rm{m}}^{2}}$ is a shift in the equilibrium position of the non-localised particle due to the force of gravity, $g$.} in the direction of gravity is given by $\Delta Z=2\frac{g_{\rm{nv}}\mu_{\rm{B}} B_{\rm{}z}}{m\Omega_m^{2}}$ which tends to be much smaller than the physical size of typical nanodiamonds~\cite{scala_matter-wave_2013,yin_large_2013}. This makes levitated sensing schemes unfriendly for resolving gravity gradients without the use of surveying or arrays.

\subsubsection{Free-fall measurements}
\label{sec:freefall}
Releasing the nanoparticle from the trap, such that it undergoes free-fall, allows its wave-function to evolve, increasing the position spread linearly in time. The ability to perform such a drop-test is unique to the levitated platform amongst optomechanical systems. Free-fall acceleration sensors are known as absolute gravimeters because they give a direct measure of gravity in units of ms$^{-2}$ traceable to metrological standards. Relative gravimeters are masses supported by a spring, for example, the stiffness of a cantilever or the optical trapping of a nanosphere. One must calibrate relative gravimeters by measuring the stiffness of the spring and placing the instrument in a location with a known gravitational acceleration. Absolute gravimeters are therefore required to calibrate relative ones. 

Perhaps the most recognisable free-fall quantum measurement is the double slit experiment, with a particle falling through a diffractive element. This can be recreated with a nanoparticle launched ballistically, with two additional cavities positioned vertically along the free-fall path. The first is used to apply a laser pulse at time $T$ that measures the square of the position via a homodyne measurement to create a vertical spatial superposition. After the superposition evolves for a further $T$ seconds, the second cavity is used to measure the particle's c.o.m. position such that after many iterations, an interference pattern is formed. The effect of gravity is to introduce a vertical shifting of the entire interference pattern on the screen by $\delta y=\frac{g}{2}(2T)^{2}$. If one arm of the superposition experiences a differing force to the other, i.e. due to a slight difference in the gravitational field, this also has a shifting effect on the \hl{interference} pattern~\cite{romero-isart_large_2011,rasel_2012}. However, creating a clear interference pattern requires many repeated launches of the same nanoparticle, requiring capture and launch techniques more mature than currently achieved, discussed in Section~\ref{sec:sen-comm}. 

For detecting transverse accelerations, i.e. to detect a nearby object placed perpendicular to the force of gravity, one can use a Talbot interferometer scheme proposed in~\cite{Geraci2015}. 

In a Talbot interferometer, a light pulse grating is applied to a free-falling quantum nanosphere, causing the nanosphere to diffract and interfere with itself. This creates an image of the grating at a distance defined by the Talbot length, and at every integer of the Talbot length. At half the Talbot length the interference pattern is phase shifted by half a period. By positioning an object adjacent to the free-fall path at a distance $<10\,\mu$m, one can probe the gravitational force produced by the object based on the transverse shift of the interference pattern. A sensitivity of $10^{-8}$\,ms$^{-2}$ is predicted for a sphere 13\,nm in diameter where a phase difference of $\Delta\phi=\pi$ corresponds to an acceleration of $4\times 10^{-6}\,$ms$^{-2}$. The fringe pattern shift is given by $\delta x_{\phi}=-a\,T^{2}_{t}$ where $a$ is the transverse acceleration and $2T_{t}$ is the total flight time with $T_{t}$ the Talbot time. 

Alternatively, one can avoid using matter interferometry, and instead, directly measure the shift in the nanosphere position caused by transverse acceleration $\delta x=\frac{at^{2}}{2}$. This can be achieved by dropping the nanosphere ballistically, such that after $t$ seconds, it falls into a cavity which then measures its shifted position. The sensitivity obtained for this ballistic scheme compared to the interferometric measurement scales with $\frac{\chi \sigma_{\rm{v}}t}{d}$, where a decrease in $\chi$, the fringe contrast, or an increase in $d$, the grating period, will reduce the advantage of the Talbot scheme over the ballistic one. The spread of the position in the ballistic set-up, given by $\sigma_{\rm{v}}t=\sqrt{\frac{\hbar\Omega_{\rm{m}}}{2m}}t$, grows during the free-fall time, contributing to the uncertainty in $\delta x$. This is another reason why the Talbot scheme is more sensitive.  The initial momentum spread due to zero-point fluctuations does not influence the position of the interference fringes in a Talbot interferometer, instead, only affecting the envelope. 

It is not straightforward to conclude when the Talbot interference scheme surpasses the ballistic scheme if parameters such as mass are varied, since this can result in a reduction of the time-of-flight due to decoherence. For example, ground state cooling of a 200\,nm diameter nanosphere enables a ballistic sensitivity 10 times higher than a Talbot measurement on a 12\,nm diameter particle ~\cite{Geraci2015}.

Lastly, the levitated spin-oscillator Ramsey interferometer scheme shown in Figure~\ref{fig:gravimetry}A can be modified for free-fall evolution, as shown in Figure~\ref{fig:gravimetry}B. Due to the long coherence time of spin states, the superposition persists even if the oscillator does not remain in a pure coherent state. The scale of the superposition is controllable through flight time and/or magnetic field gradient such that the acquired phase is given by \cite{wan_free_2016}: 
\begin{equation}
    \Delta\phi=\frac{1}{16\hbar}g_{\rm{nv}}\mu_{\rm{B}}B_{\rm{x}}g t_{3}^{3}\cos(\theta),
    \label{eq:freefall}
\end{equation}
where $g_{\rm{nv}}$, $\mu_{\rm{B}}$ are defined as above, $\theta$ is the angle between the applied magnetic field $B_{\rm{x}}$ and the direction of the gravitational acceleration $g$ and $t_{3}$ is the total free-fall time (i.e. when the wavepackets merge and interfere).

In contrast to the levitated Ramsey scheme, equation~\ref{eq:freefall} does not depend on the mass, \hl{as expected for a free-fall measurement. An} additional microwave pulse is required at time $t_{3}$ to reverse the propagation direction of the superposed wave packets. This pulse flips the spin state such that the spin dependent force will reverse. The split wavepackets eventually merge and interfere. The measurement time $t_{3}$ is therefore unconstrained and can be on the order of milliseconds, enabling spatial superpositions spanning~100\,nm, over $10^{3}$ times larger than if the nanoparticle was levitated \cite{wan_free_2016}, and comparable in scale to the size of the particle. The maximum spatial separation along the \emph{tilted} $x$-axis is $\Delta x_{\theta}=2\frac{g_{\rm{nv}}\mu_{\rm{B}}B_{\rm{x}}}{2m}(t_{3}/4)^{2}$, at $t=t_{3}/2$. Currently, coherent scattering cooling would achieve a maximum coherence time of 1.4\,$\mu s$ in free-fall, limited by background pressure, which only allows for an expansion of the wavepacket from 3.1\,pm to 10.2\,pm~\cite{delic_cooling_2020}. Further progress will require deep vacuum environments and likely cryogenic operating conditions.

\subsection{Comparison}
Table~\ref{tab:comparison} shows a comparison of the predicted and achieved acceleration sensitivities obtained by quantum and classical research sensors, alongside the current commercial state-of-the-art. \hl{Also included is the achieved accuracy obtained by two sensors. Note that sensitivity, defined by the velocity random walk on an Allan deviation plot of the sensor output~\footnote{\hl{For pulsed measurements, common in atom interferometry, sensitivity does not imply a spectral density.}}, is a measure of precision, but does not guarantee accuracy which describes the trueness of the value. Academic devices that are not based on cold atoms require calibration, and therefore not as easily traceable to the International System of Units (SI).} We focus on devices which are suitable for gravimetry but would need to be used in a flywheel operation with a classical inertial measurement unit (IMU) for navigation applications. The latter requires sampling rates above 100\,Hz, incompatible with the time of flight used in free-fall experiments or the pulse sequence needed for Ramsey interferometry. Flywheel operation uses a classical IMU to provide inertial measurements in-between this deadtime, and is used by cold atom inertial sensor prototypes~\cite{battelier_development_2016}. In turn, the quantum measurement, which is less susceptible to drift, is used to reset the growing errors accrued by the IMU. The achievable sensitivities across all types of sensors are comparable, which is unsurprising considering the majority are operated classically, where optimisation of the detection noise and/or effective mass can still significantly improve sensitivity at the cost of bandwidth. However, all current sensors struggle to surpass a sensitivity better than $10^{-9}\,$ms$^{-2}$.

An interesting question is: what sets the fundamental limit in sensitivity at the quantum level? The field of quantum metrology seeks to find these fundamental limits through use of quantum Fisher information (QFI) which is a metric of intrinsic accuracy \hl{(assuming all unknown variables can be traced to SI units)}, and only depends on the input state. It is formally defined via the Cram{\'e}r-Rao bound as the inverse of the variance of a measurable property, in this case, phase. A high QFI ensures greater precision. The QFI cannot reveal the underlying measurement protocol to obtain such limits, but one may test a measurement protocol by computing its associated classical Fisher information (CFI). The CFI takes into account both the input state and the extractable information from the measurement scheme, and may or may not meet the QFI bound~\cite{armata_quantum_2017,schneiter_optimal_2020}.

\begin{table}
\begin{tabular}{>{\arraybackslash}m{2.8cm}|>{\arraybackslash}m{5.1cm}}
\hl{Existing} System & Achieved sensitivity (ms$^{-2}\,$Hz$^{-1/2}$)\\
\hline
Free-fall cube mirror$^{\dagger}$ &  $1.5\times 10^{-7}$ \textmd{($10^{-9}$ms$^{-2}$ in 6.25hr)}\cite{lacoste}$^{\ddag}$ \\
Atom interferometer & $4.2\times 10^{-8}$ \textmd{$(3\times 10^{-9}$ms$^{-2}$ in 300s)}\cite{hu_2013}\\
\hl{Atom interferometer} & \hl{$9.6\times 10^{-8}$ \textmd{$(5\times 10^{-10}$ms$^{-2}$ in 2.8hr)}{\cite{Freier_2016}}}\\
Lev. optomechanics$^{\dagger}$ & $4\times 10^{-6}$ \textmd{($6.9\times 10^{-9}$ms$^{-2}$ in 3.8hr)}\cite{monteiro_optical_2017} \\
Lev. optomechanics$^{\dagger}$ & $9.3\times 10^{-7}$ \cite{monteiro_force_2020}$^{\clubsuit}$\\
Opto-MEMS$^{\dagger}$ & $7.8\times 10^{-8}$ \cite{krishnamoorthy_-plane_2008}\\
Capacitive-MEMS$^{\dagger}$ & $3\times 10^{-9}$ \cite{pike_2019}\\
\hline
\hl{Existing System} & \hl{Achieved accuracy (ms$^{-2}$)}\\
\hline
\hl{Free-fall cube mirror$^{\dagger}$}  & $2\times 10^{-8}$~\cite{lacoste}\\
\hl{Atom interferometer} & \hl{$3.9\times 10^{-8}${~\cite{Freier_2016}}}\\
\hline
 &\\
\hline
\hl{Proposed System} & \hl{Predicted} sensitivity (ms$^{-2}\,$Hz$^{-1/2}$) \\
\hline
Trapped cold atom* & $\times 10^{-10}$ \\
Lev. optomechanics* & $\times 10^{-15}$ \\
Lev. spin-mechanics & $2.2\times 10^{-9}$ \cite{johnsson_macroscopic_2016} \\
Talbot optomechanics & $\times 10^{-8}$ \cite{Geraci2015} \\
\hline
\end{tabular}
\caption{\label{tab:comparison}Table showing the achieved and predicted acceleration sensing sensitivities among various types of accelerometer. \hl{We include two measurements of sensor accuracy. Optomechanical systems require further work to define their accuracy as they are only calibrated against commercial sensors.} 'lev.' denotes levitated experiments, BEC is a Bose Einstein Condensate, and MEMS are micro-electro-mechanical systems. $^{\dagger}$ denotes classical measurements, noting that capacitive MEMS do not currently operate in the quantum regime and optomechanical devices have only recently entered the quantum regime. $^{\ddag}$ is the value from LaCoste's brochure whereas $5.6\times 10^{-7}$\,ms$^{-2}$\,Hz$^{-1/2}$ was measured in the laboratory~\cite{wang_shift_2018}. $^{\clubsuit}$ indicates the sensitivity whilst the c.o.m. temperature of the oscillator is cooled to $50\mu$K~\cite{monteiro_force_2020}. * is the predicted sensitivity using classical Fisher information assuming a homodyne detection scheme, obtaining a sensitivity close to that predicted by quantum Fisher information \hl{{\cite{Qvarfort2018}}}.}
\end{table}

In \hl{{\cite{Qvarfort2018}}} the CFI for a homodyne measurement of a quantum levitated nanosphere using continuous optomechanical transduction is calculated, which they compare with the CFI for a typical atom interferometry set-up. Over five orders of magnitude improvement in sensitivity is predicted for a levitated optomechanical oscillator versus a cloud of $10^{5}$ atoms, shown in the lower portion of Table~\ref{tab:comparison}. This clearly highlights the potential competitive advantage in quantum levitated optomechanics, including the relative ease in preparing a single macro-sized object over a dense cloud of atoms. \hl{However, we stress that such a claim requires further work to improve the accuracy of optomechanical sensors, involving precise calibration of mass and frequency to SI standards. cold atom measurements are intrinsically traceable.}

\subsection{Road to commercialisation}
\label{sec:sen-comm}
For some time the optomechanics community has been prototyping classical accelerometers \cite{li_field_2018,huang_chip-scale_2020,krause_high-resolution_2012,guzman_cervantes_high_2014,monteiro_optical_2017,monteiro_force_2020,middlemiss_field_2017}. Such sensors rarely require the level of environmental isolation needed for long-lived quantum state preparation. Once the community is able to repeatedly demonstrate quantum state preparation of levitated nanospheres, now possible with coherent scattering techniques, they should use the advancement of commercial cold atom interferometers as a roadmap for developing application-ready tools. 

Common to both quantum optomechanics and cold atom interferometry is the need for ultralow and stable vacuum pressures, preferably at~$10^{-12}$\,mbar or below, to prevent collisions with background gas molecules. Lowering the environmental temperature using cryostat technologies reduces the influence of thermal heating, although this is more crucial for clamped optomechanics experiments using cantilevers, membranes or MEMS structures\hl{. Levitation} minimizes thermal contact with the environment. For the coherent scattering set-up, they measure background collisions as their largest source of decoherence when at a pressure of $10^{-6}$\,mbar, requiring $10^{-11}$\,mbar and cryostat temperatures below 130\,K to sustain a wavepacket on the order of the particle radius~\cite{delic_cooling_2020}. The stability of the pressure, ambient temperature, and laser frequency and intensity require consideration, as these may be sources of decoherence or drift that skew or washout the measured signal. Shot-noise limited laser sources are best suited. 

Any vibrational noise in the components will also create errors, with vibrations at low frequency the most critical. A baseline vibrational stability of nm\,Hz$^{-1/2}$ is recommended~\cite{Geraci2015}, achievable with modern commercial cyrostats. Another source of mechanical error is misalignment. For example, a vertical tilt when measuring gravity will create an offset, characterised as noise if the alignment varies per shot. Tilt fluctuations no higher than $0.5\,\mu$rad\,Hz$^{-1/2}$ are recommended~\cite{Geraci2015}. 

Unique to levitated optomechanics is the need for a launch and recapture method, since nanospheres are not indistinguishable like atoms. So far, loading at low vacuum pressure has been achieved through (i) momentum imparted by either a piezo-speaker \cite{millen_cavity_2015} or a laser induced acoustic shock of a wafer covered by tethered nanorods \cite{kuhn_cavity-assisted_2015}, (ii) electrospray injection of particles \cite{bullier_characterisation_2020}, or (iii) conveyor-belt loading using optical \cite{grass_optical_2016} or electrical forces \cite{bullier_characterisation_2020}. Reliable recapture of a levitated nanosphere still remains a technical challenge, but can be built upon successful proof of principle demonstrations~\cite{hebestreit_sensing_2018}. Alternatively, sourcing nanospheres with high reproducibility enables continuous injection of particles, where an in-situ calibration could be performed to account for size variances. Nanoparticles also vary in their surface charges, requiring shielding from stray electric fields to avoid dephasing from Coulomb force interactions. Recently, it was demonstrated that levitated particles could be discharged using a high voltage wire that ionizes residual gas molecules, adding charges to environment~\cite{frimmer_controlling_2017}. Single elementary charge precision was achieved, which would enable zero net charged nanoparticles to be prepared when starting with a mixture of the number of charges~\cite{ranjit_attonewton_2015}.

Lastly, all sensors must pass certain environmental testing conditions to be deployed for space/aerospace, military and metrology use. These include high electromagnetic interference protection levels, operational temperatures between -40$^{o}$C to +85$^{o}$C and shock resistance (in some cases, up to 20,000\,ms$^{-2}$). Testing is conducted in shake and bake chambers that apply acceleration whilst cycling the ambient temperature defined by standards such as those used by the military (MIL-STD), aerospace (DO) or consumer use (CE marking in EU)~\cite{honeywell_standards}. Those who perform these tests must also obtain certification that they meet the requirements governed by the International Organization for Standardization (ISO). Reduction in size, weight and power (SWaP) alongside cost are also factors that require involvement with supply chains and industry. Through global quantum technology initiatives, such collaborations are already underway, resulting in miniature vacuum chambers and chipscale ion traps \cite{schwindt_highly_2016,rushton_contributed_2014,birkl_micro_2007} developed for cold atoms, and miniaturised quantum sources of light \cite{caspani_integrated_2017} on chip-scale photonic integrated circuits developed for quantum computing. Preliminary feasibility studies have also been carried out in partnership with the European Space Agency to mature the supporting technologies needed to implement macroscopic state preparation and interferometry using levitated nanospheres in space~\cite{kaltenbaek_macroscopic_2016}.

\section{Outlook}

In this viewpoint we have reviewed proposals that aim to implement acceleration sensing using spatial superpositions of quantum levitated nanoparticles. Performing macroscopic matter-wave interferometry is now tantalisingly close with the advent of coherent scattering, used to cool a 143\,nm diameter sphere to the ground state of an optical potential \cite{delic_cooling_2020}. A new regime of quantum levitated optomechanics is upon us, 10 years after the first \emph{clamped} human-made object was cooled to the ground state~\cite{oconnell_quantum_2010,aspelmeyer_surf_2010}. The potential advantage of using a macroscopic quantum test-mass has been theoretically predicted using quantum and classical Fisher information analysis; 5-orders of magnitude improved sensitivity is expected over existing cold atom sensors. Although such predictions do not reflect technical feasibility, a potential demonstration of second-generation quantum sensing using optomechanics is certainly on the horizon. 

Similar to cold atoms, there are a variety of quantum sensing protocols proposed in levitated optomechanics, whereby the oscillator either remains trapped or undergoes free-fall to enable larger spatial superpositions and averaging times. Many challenges for field-testing free-fall quantum sensors are already being solved by the commercial development of cold atom interferometers, backed by significant funding from governments and industry across the world~\cite{gibney_quantum_2019}. Unique to levitated optomechanics are challenges in reproducible and reliable launch, capture, and characterisation of nanoparticles with size variations, or the fabrication of near-identical nanoparticles. 

At the time of writing, the UK, the EU, the US, China, Russia and Canada have or will be committing over £1B each to their respective quantum technology initiatives~\cite{thew_focus_2019}. Although quantum optomechanical sensors may not mature at the same rate as other quantum technologies, progress is undoubtedly linked to the successful commercialisation of existing cold atom sensors or quantum communication devices. These disruptive technologies share common goals, similar routes to market, and interchangeable subcomponents. We call upon the wider scientific community for increased cross pollination of resources and methods, wider engagement with industry, and global collaborations.

\begin{acknowledgement}
  The authors thank Antonio Pontin for discussions. \hl{The authors also thank Dennis Schlippert and Christian Schubert for their valuable comments and discussions on Table~\ref{tab:comparison} and Anupam Mazumdar for their feedback on quantum gravity}.
\end{acknowledgement}
\begin{funding}
YLL is supported by a Royal Academy of Engineering Intelligence Community Postdoctoral Fellowship Award: ICRF1920\textbackslash3\textbackslash10. MR acknowledges funding from the EPSRC Grant No. EP/S000267/1. JM is supported by the European Research Council (ERC) under the European Union's Horizon 2020 research and innovation programme (Grant Agreement No. 803277), and by EPSRC New Investigator Award EP/S004777/1.
\end{funding}
\nocite{*}
\printbibliography[category=cited,heading=subbibliography]
\end{document}